# Elucidating the initial steps in α–uranium hydriding using first-principles calculations

*By Artem Soshnikov,[1] Ambarish Kulkarni,[1] Nir Goldman[1,2]*

[1]Department of Chemical Engineering, University of California, Davis, CA 95616, USA

[2]Physical and Life Sciences Directorate, Lawrence Livermore National Laboratory, Livermore, CA, 94550 USA.

Email: goldman14@llnl.gov

# ABSTRACT

Hydrogen embrittlement of uranium, which arises due to the formation of a structurally weak pyrophoric hydride, poses a major safety risk in material applications. Previous experiments have shown that hydriding begins on top or near the surface (i.e., subsurface) of $\alpha$-uranium. However, the fundamental molecular-level mechanism of this process remains unknown. In this work, starting from pristine $\alpha$-U bulk and surfaces, we present a systematic investigation of possible mechanisms for formation of the metal hydride. Specifically, we address this problem by examining the individual steps of hydrogen embrittlement, including surface adsorption, subsurface absorption, and the inter-layer diffusion of atomic hydrogen. Furthermore, by examining these processes across different facets, we highlight the importance of both (1) hydrogen monolayer coverage and (2) applied tensile strain on hydriding kinetics. Taken together, by studying previously overlooked phenomena, this study provides foundational insights into the initial steps of this overall complex process. We anticipate that this work will guide near-term future development of multiscale kinetic models for uranium hydriding and subsequently, identify potential strategies to mitigate this undesired process.

**INTRODUCTION**

Uranium is a unique element that can be used in the energy industry as a nuclear fuel to generate electricity, and by the military to power submarines and for weaponry.[1] However, due to the high inherent reactivity of the metal, even trace amounts of hydrogen gas can readily induce corrosion. Under ambient conditions, uranium and hydrogen spontaneously combine to form uranium hydride ($UH_3$), resulting in the physical disintegration of the parent metal. There are two known $UH_3$ phases: the $\alpha$-$UH_3$ phase has a higher symmetry and typically observed at low temperature (<80 °C), while the $\beta$-$UH_3$ has a lower symmetry and more prominent at high temperature (>200 °C).[2] As uranium incorporates hydrogen, the uranium lattice expands by approximately 75% in volume, causing the formation of a black dispersive powder under ambient conditions with high surface area.[2] Uranium hydride is highly toxic and pyrophoric, releasing enormous amounts of heat upon exposure to air.[3] To date, there has been very limited understanding of how to control or mitigate the embrittlement process. Specifically, an atomic-level understanding of hydrogen-induced corrosion could yield enhanced safety regulation policies and mitigation of toxic waste.

In the past few decades, many studies have investigated the thermochemistry, permeability, and diffusion of atmospheric gases within uranium. Experiments have shown that the hydriding process is characterized by the appearance of growing "spots," or surface monolayers.[4] These monolayer sites expand radially and eventually merge to form a continuous layer of hydride on the metal surface. Studies by Mallet and Trzeciak at 1 atm and 537 K showed the solubility of hydrogen in $\alpha$-U to be 9.3x10$^{-5}$ H atoms per unit cell U,[5] far below the U/H ratio of 1:3 in $\alpha/\beta$-$UH_3$. In another study by Powell and Condon (reported by Condon and Larson), a diffusional barrier of 0.502 eV was estimated by tracking the preloaded hydrogen degassing rate from a uranium foils and determining an Arrhenius relationship for the diffusion constant.[6] These values would appear to indicate that the hydriding processes exhibits significant barriers under ambient conditions, in stark contrast to the usual spontaneous process. Previous experiments have largely investigated bulk properties[2] as well as hydrogen attacking regions beneath the hydride craters,[7] and did not directly probe surface effects. Such studies could overlook the potential importance of trap sites and the prevalence of hydriding at the surface and near the subsurface. In addition, hydriding experiments on pure uranium are difficult under ambient conditions. The uranium surface is typically covered with a protective oxide layer when exposed to air, which acts as a barrier to hydrogen reactivity, including diffusion and dissociation, and introduces a factor of unpredictability into the observed hydriding induction times. These studies would greatly benefit from atomic-level detail about hydrogen-uranium surface reactions, nucleation, and growth parameters as hydrided layers begin to form.

Density Functional Theory (DFT) is a well-established computational approach in material science, physics, and chemistry for the prediction of physical and chemical properties. DFT has been used to study numerous metal-hydrogen systems across the periodic table,[8–13] including palladium alloys,[14] titanium,[15] and many other systems. It has also been used to compute bulk absorption in plutonium[16] as well as hydrogen monolayer coverage on its low energy facets.[17] Several previous studies have probed the initial interactions of α-U and hydrogen in the dilute limit (i.e., a single hydrogen atom per surface) using DFT. For example, some results exist regarding surface energies, single atom/molecule geometries, point defect formation energies in the bulk, and adsorption energies on the (001) surface (generally considered to be the most stable).[18–23] While these studies provide useful information about the very first steps of the hydriding process (including $H_2$ dissociation), the effects of strain, concentration and partial pressure of hydrogen, and monolayer coverage remain entirely unknown. Hence, our effort is the most in-depth ab initio study of U-H interactions as well as the uranium hydriding phase diagram that we know of to date.

In this paper, we greatly expand upon these results by using DFT to further examine possible mechanisms for uranium hydriding, including bulk effects, the presence of multiple surfaces, and a wide range of hydrogen concentrations. First, we perform a thorough analysis of DFT exchange-correlation functionals and theoretical approaches in order to determine an appropriate computational protocol that reproduces experimentally known bulk and surface properties. Hydrogen embrittlement and the phase transformation to α/β-$UH_3$ involve both changes in chemical composition and volumetric expansion of the lattice. Consequently, we investigate the process of uranium hydriding from two perspectives: (1) from formation of hydrogen monolayer coverage and (2) as a function of applied tensile load on the α-U lattice. We believe that both of these mechanisms could play significant roles in hydrogen embrittlement, and we quantify the relative importance of different surface adsorption sites and facets in the process.

**COMPUTATIONAL DETAILS**

Density Functional Theory (DFT) calculations were performed using the Vienna ab initio simulation package code[24] (VASP). The projector augmented wave[25] (PAW) pseudopotentials for U provided in VASP include $6s^2 6p^6 5f^3 6d^1 7s^2$ as valence electrons. Fourth order Methfessel-Paxton smearing[26] was used with a value of 0.2 eV for all optimizations calculations in order to ensure energy convergence without dependence on the electronic smearing temperature. In our initial calculations, we examined the following exchange-correlation functionals: Perdew-Burke-Ernzerhof (PBE), dispersion corrected PBE (PBE_D3BJ, RPBE_D3BJ), and PBEsol and SCAN, which were developed specifically for solids and solid surfaces. The energy cutoff for the planewave basis set for all of our calculations was set to 500 eV based on convergence tests. Structural relaxations were performed until forces on each atom were less than 0.01 eV/Å. A k-point mesh of 14x7x8 generated by the MonkhorstPack[27] method for integration over the Brillouin zone was used for the primitive bulk unit cell optimization.

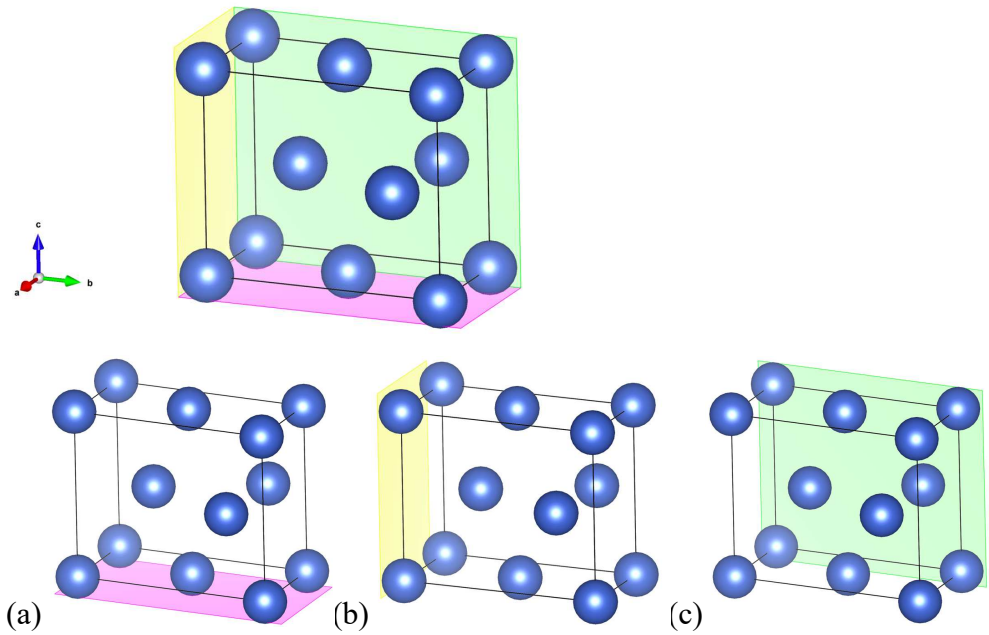

**Fig. 1:** The α-U surface unit cell (a) (001) – magenta, (b) (010) – yellow, and (c) (100) – green.

The alpha phase of uranium has a face-centered orthorhombic structure with a CmCm space group[28], as shown in Fig. 1. To investigate the interaction between atomic hydrogen and α-U surfaces, a slab model with at least a six-layer thickness was employed, in which the bottom two layers were constrained to simulate the bulk environment while the rest were allowed to relax their equilibrium positions. A vacuum layer of 15 Å between two adjacent slab surfaces was found to be adequate for all relevant calculations.

The surface adsorption and subsurface absorption behavior of hydrogen was modeled using a (2x1) surface unit cell with 6-10 layers depending on the facet (between 24-40 U atoms), while the bulk absorption was modeled using a 4x2x2 supercell (64 U atoms). For a converged, clean slab, the surface energy was calculated using the following formula:

$$E_{surf} = \frac{E_{slab} - N * E_{bulk}}{2A}, \qquad (1)$$

where $E_{slab}$ is the total energy of the N-atom slab, $E_{bulk}$ is the bulk energy per atom, A is the area of the surface, and N is the number of atoms in the surface slab. $E_{bulk}$ is approximated using Fiorentini and Methfessel[29] approach from the slope of a linear fit of DFT energies from slabs of different thickness.

The adsorption energy of one H atom on an α-U surface and the absorption energy into a subsurface site were determined using the following expression:

$$E_{ads/abs} = E_{(U/H)} - E_{(U)} - {}^1\!/_2\, E_{(H_2)}, \qquad (2)$$

where $E_{(U/H)}$ is the total energy of the optimized hydrogen+slab system, $E_{(U)}$ is the total energy of the clean slab, and $E_{(H_2)}$ is the total energy of the optimized isolated $H_2$ molecule. A negative adsorption energy corresponds to a stable adsorbate/surface system.

**RESULTS & DISCUSSION**

**I. FUNCTIONAL COMPARISON**

**A. Lattice parameters**

Before we investigate uranium surface chemistry, we first perform a benchmark comparison study of the equilibrium lattice parameters with various exchange-correlation functionals with experimental values. As shown in Table 1, all exchange-correlation functionals used here underestimate the experimentally reported lattice constants. However, these deviations are relatively small (less than 3.97%). Even though PBEsol is designed to improve upon PBE for equilibrium properties of closely packed solids, it slightly underperforms its predecessor in terms of estimating lattice parameters. In addition, the long-range dispersion effects in PBE_D3BJ and RPBE_D3BJ also do not yield improved results, as these functionals display the worst calculated lattice constants. SCAN, which is a meta-GGA functional and thus includes the second derivative of the electron density, shows better results than all D3 functionals and PBEsol, but slightly worse than PBE. Overall, we observe that PBE and SCAN showed the lattice constants slightly closer to the experimental values, followed by PBEsol, PBE-D3BJ, and RPBE-D3BJ.

We note that standard DFT functionals tend to underestimate electron correlations for 5f materials (e.g., U, Pu), which in turn can result in an underestimation of the atomic volume.[30] In general, this effect can be corrected through the inclusion of semi-empirical on-site Coulombic repulsions.[31] In order to test this effect on the bulk, we have performed GGA+U calculations using a $U_{eff} = U - J$ value of 1.02 eV which was computed from linear response theory.[32] As shown in Table 1, the GGA+$U$ approach somewhat improves α-U lattice parameter prediction, though its effects are relatively small, similar to previous results.[33]

**TABLE 1**: Estimated equilibrium lattice parameters for α-U unit cell using PBE, PBEsol, PBE_D3BJ, RPBE_D3BJ, and SCAN functionals. Percent deviation from the published experimental result are indicated in parenthesis.

| method | a (Å) | b (Å) | c (Å) | volume (Å³) |
|---|---|---|---|---|
| PBE | 2.811 (-1.16%) | 5.857 (-0.20%) | 4.901 (-0.63%) | 80.690 (-1.98%) |
| PBEsol | 2.745 (-3.48%) | 5.753 (-1.98%) | 4.839 (-1.89%) | 76.417 (-7.17%) |
| PBE_D3BJ | 2.752 (-3.23%) | 5.729 (-2.39%) | 4.853 (-1.60%) | 76.513 (-7.06%) |
| RPBE-D3BJ | 2.731 (-3.97%) | 5.681 (-3.20%) | 4.830 (-2.07%) | 74.937 (-8.97%) |
| SCAN | 2.803 (-1.44%) | 5.854 (-0.26%) | 4.898 (-0.69%) | 80.370 (-2.37%) |
| PBE+$U$ | 2.851 (+0.25%) | 5.864 (-0.09%) | 4.957 (+0.51%) | 82.872 (+0.67%) |
| EXP[34] | 2.844 | 5.869 | 4.932 | 82.322 |

## B. Surface energies and atomic hydrogen adsorption on α-U (001) surface

After optimizing and comparing the lattice constants, we also performed an exchange-correlation functional comparison study by computing surface energies (Table 2). Here, we chose the (001) surface as this is the most stable α-U facet. The experimental surface energy of α-U is 1.828 J/m$^2$, however, this value corresponds to an isotropic crystalline structure of the metal or an average value of different surfaces.[19] Since the (001) facet has the lowest energy, it would be the most abundant surface in uranium crystalline material. As presented in Table 2, the descending order of the surface energy for various functionals is RPBE_D3BJ > PBE_D3BJ > PBEsol > PBE > SCAN. The experimental surface energy is in close proximity to the results determined from PBE and SCAN, while estimates by other functionals are much higher.

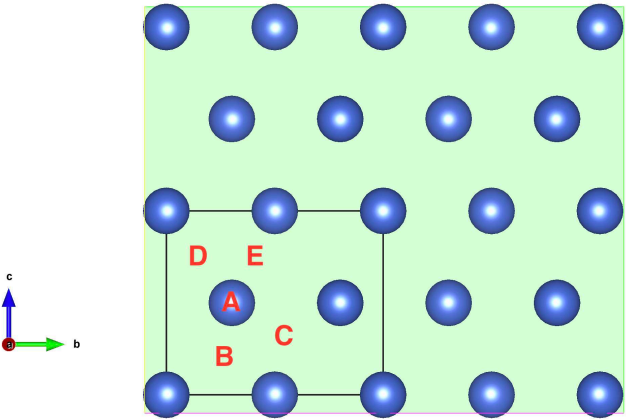

**Fig. 2**: Schematic illustration of initial adsorption sites on (001) surface. A, B, C, D and E represent the initial sites of top, hollow1, hollow2, longbridge, and shortbridge, respectively.

**TABLE 2**: Energetic parameters and geometric features of H atoms on α-U (001) surface estimated using PBE, PBEsol, PBE_D3BJ, RPBE_D3BJ, and SCAN functionals. $E_{ads}$ denotes the adsorption energy, $d_{H-surf}$ represents the average binding height from H atom to the top layer of the slab, and SE denotes the surface energy for the pristine surface.

| Approach | Ads Site | $E_{ads}$ (eV) | $E_{ads}$ (eV) w/zpe | $d_{H-surf}$ (Å) | SE (J/m²) |
|---|---|---|---|---|---|
| PBE | hollow1 | -0.51 | -0.35 | 2.229 | 1.839 |
|  | hollow2 | -0.57 | -0.41 | 2.221 |  |
| PBEsol | hollow1 | -0.59 | -0.43 | 2.226 | 2.191 |
|  | hollow2 | -0.65 | -0.49 | 2.219 |  |
| PBE_D3BJ | hollow1 | -0.58 | -0.42 | 2.231 | 2.834 |
|  | hollow2 | -0.64 | -0.48 | 2.225 |  |
| RPBE_D3BJ | hollow1 | -0.58 | -0.42 | 2.223 | 3.482 |
|  | hollow2 | -0.62 | -0.46 | 2.215 |  |
| SCAN | hollow1 | -0.51 | -0.35 | 2.274 | 1.832 |
|  | hollow2 | -0.57 | -0.41 | 2.264 |  |

Van der Waals (vdW) interactions are generally significant for surface properties and their inclusion in DFT models remains an area of active research.[35] PBE generally neglects most of the vdW interactions, while SCAN incorporates intermediate-range vdW, and D3 functionals include even long-range dispersion effects. Remarkably, these dispersion effects are of less importance in estimating bulk and surface properties in α-U. As mentioned, the closest agreement in estimating the surface energies was achieved by PBE and SCAN. SCAN is considered a more advanced functional due to inclusion of all 17 known exact constraints that a meta-GGA can satisfy. However, we only notice slight improvements in results calculated by SCAN over PBE.

In addition to surface energies, we performed a similar comparison study in estimating atomic hydrogen adsorption energies on the most stable (001) α-U surface (Fig. 2). As shown, five possible unique adsorption sites (denoted as top, hollow1, hollow2, longbridge, or shortbridge) were considered. For all cases, the H atom is unable to remain stable at the top, long-bridge, or shortbridge sites, relocating to a more stable hollow site nearby. This indicates that there are only two preferred adsorption sites on the (001) surface for a single hydrogen adsorption: hollow1 and hollow2. Similar results are observed in nitrogen/uranium[36] and oxygen/uranium[37] systems. In addition, all functionals used here indicate that the hollow2 site is slightly more energetically favorable (~0.06 eV) compared to hollow1. Table 2 also shows minimal differences between functionals, with a relatively small energetic spread of ~0.08 eV for all approaches, due in part to the various representations of vdW interactions. We note that quantum nuclear vibrational zero-point energies (ZPE) can be significant for low-Z elements such as hydrogen. We observe that the ZPE occurs as a systematic correction of ~0.16 eV for all functionals studied here. We discuss the effects of ZPE further regarding reaction energies and kinetics (section IV) and our computed monolayer coverage phase diagrams (Section IV).

Overall, we have computed the lattice constants, surface energies, and atomic hydrogen adsorption energies on the (001) facet. We have compared experimental determined results to those calculated by various exchange-correlation functional approaches. Given the correct ordering and reasonably small differences in adsorption energies, surface energies, and lattice constants results, PBE was our functional of choice for the rest of our studies.

Surface energies and adsorptions have been shown to be relatively insensitive to the value of the Hubbard U correction in actinides.[38] Similar results have been seen with materials containing lower-Z metals, where inclusion of GGA+U had a minimal effect on surface energy ordering.[39] In addition, uranium phases are known to exhibit magnetic moments at their surface,[40,41,42] which can affect surface adsorption energies. In order assess these effects on α-U surface properties and hydrogen adsorption, we have chosen to perform a bounding study using PBE on the faceted (012) and (102) surfaces, which are the two highest energy/most reactive surfaces studied in our work (see Section III for further discussion). Here, we compute surface energies and hydrogen adsorption energies for the lowest energy site (i.e., the stepped site) using GGA+U and spin polarization separately in order to independently quantify the effect of each treatment. This approach places an upper bound on these effects since these surfaces likely exhibit the largest surface magnetization or other sensitivities due to choice of DFT method.

**TABLE 3**: Comparison of surface properties of the (012) and (102) facets computed by different methods. The signifier 'SP' denotes spin polarization.

| Surface | PBE (non-SP) | PBE (SP) | | PBE+U (non-SP) |
|---|---|---|---|---|
| | SE (J/m$^2$) | SE (J/m$^2$) | Mag mom/surf atom | SE (J/m$^2$) |
| (012) | 2.15 | 2.05 | 2.23 | 2.07 |
| (102) | 1.92 | 2.03 | 1.94 | 2.02 |

**TABLE 4**: Comparison of optimal hydrogen adsorption energies on the (012) and (102) facets computed by different methods. The signifier 'SP' denotes spin polarization.

| Surface | PBE (non-SP) | PBE (SP) | PBE+U (non-SP) |
|---|---|---|---|
| (012) | -0.71 | -0.62 | -0.84 |
| (102) | -0.66 | -0.59 | -0.64 |

In the case of our GGA+U calculations, we find that the resulting surface energies experience small changes for each facet (Table 3), with the (012) surface shifting to a lower value by ~4% (2.15 J/m$^2$ vs. 2.07 J/m$^2$ for GGA+U) and the (102) surface shifting upwards by ~5% (1.92 J/m$^2$ vs. 2.02 J/m$^2$ for GGA+U). Additionally, we find that this shift results in a fairly small change in hydrogen adsorption energy (Table 4), where adsorption is enhanced (e.g., more negative) by 0.13 eV on the (012) facet and is diminished (more positive) by 0.02 eV on the (102) facet. We note that these deviations are similar in magnitude to the spread of adsorption energy results due to choice of exchange-correlation functional, shown in Table 2 of the main manuscript. Hence, we have opted not to include GGA+U corrections, given that these deviations represent a relatively small upper bound on our calculations.

Regarding spin polarization, we observe significant spin moments on the surface uranium atoms, with values of ~2 for each surface (Table 3). However, the effect on the computed surface energies is relatively small, with the (012) surface energy shifting to a lower value by ~5% (down to 2.05 J/m$^2$) and the (102) surface energy shifting higher by ~6% (up to 2.03 J/m$^2$). Similar to GGA+U, the effect on hydrogen adsorption energies is relatively small (Table 4). We find that (012) adsorption is diminished by 0.09 eV and that (102) adsorption is diminished by 0.07 eV. As a result, we find no substantial loss in accuracy due to excluding spin polarization from the results presented here as well. A systematic investigation to further quantify GGA+U and spin polarization effects on uranium surface properties is the subject of future work.

## II. ABSORPTION INTO α-U (001) FIRST SUBSURFACE LAYER

Upon molecular dissociation, atomic hydrogen will tend to bind in thermodynamically favored surface adsorption sites. This can be followed by diffusion of hydrogen from the surface to the subsurface, which could also play a substantial part in hydrogen embrittlement process. We now calculate the energetic parameters of hydrogen absorption into the first subsurface layer. In this section, we present results for the (001) facet, while results (for the most stable site) for all other surfaces of interest are presented in Table S2 (Supplemental Information; SI).

Fig. 3 shows the typical interstitial sites for the (001) surface: tetrahedral (Tet), square-pyramidal (Sqpy) and octahedral (Oct). Due to the low symmetry of the α-U surface, there are one Oct and two types of Tet and Sqpy sites between the top and the first subsurface layers. Tet1 has three surface and one subsurface uranium nearest neighbors, while Tet2 is surrounded by one surface and three subsurface U atoms (Fig. 3a, b). Sqpy1 has three surface and two subsurface U nearest neighbors, while Sqpy2 has two surface and three subsurface U atoms (Fig. 3c, d). Oct has three surface and three subsurface U atom nearest neighbors (Fig. 3e).

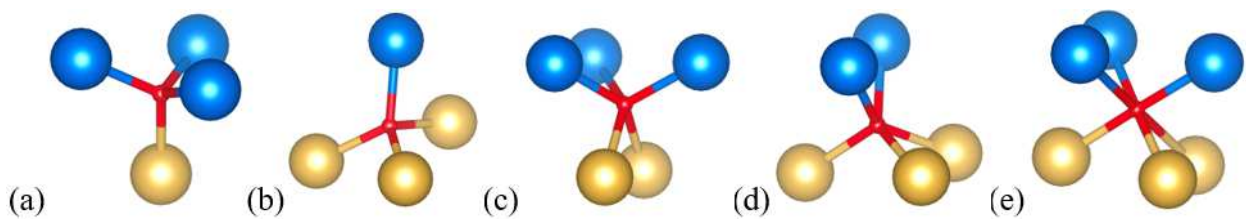

**Fig. 3**: High symmetry interstitial sites of hydrogen atoms in the first subsurface layer of α-U (001) surface: (a) Tet1 (b) Tet2 (c) Sqpy1 (d) Sqpy2 (e) Oct. The top layer U atoms are labeled in blue, the second layer U atoms in yellow, and the hydrogen interstitial in red.

The subsurface interstitial/absorption energies for each unique site were calculated using Eq. 2. The absorption energies, site type, the distances between absorbed hydrogen and top surface layer are listed in Table 5. All of the subsurface absorptions in the first sublayer were found to be endothermic. The subsurface absorption at Tet1 site was found to be the most favorable (0.16 eV), followed by Sqpy1 (0.25 eV), Sqpy2 (0.32 eV), and Tet2 (0.49 eV). In fact, the Tet1 value is almost half of the absorption energy of 0.32 eV at tetrahedral interstitial site in bulk (as per our own calculations). First layer subsurface absorption at the Oct site was found to be unstable as hydrogen atom prefers to migrate to the more stable Sqpy1 site during geometry optimization. We notice a trend of sites closer to the surface having lower energetics, in correlation with lower $d_{H-surf}$ values, shown in Table 5.

**TABLE 5**: The subsurface absorption energies and geometric parameters of hydrogen atoms under (2x1) α-U (001) surface. $E_{abs}$ denotes the average absorption energy to insert H atom inside the first subsurface layer; $d_{H-surf}$ represents the distance between H atom to the top surface layer.

| Interstitial Type | $E_{abs}$ (eV) | $d_{H-surf}$ (Å) |
|---|---|---|
| Tet1 | 0.16 | 0.463 |
| Tet2 | 0.49 | 1.884 |
| Sqpy1 | 0.25 | 0.818 |
| Sqpy2 | 0.32 | 1.379 |
| Oct | ---- | ---- |

We performed similar analysis of subsurface absorption on other low-index facets and the most favorable energetic sites are shown in Table S2 (Supplemental Information). In general, we found that these are either of Tet or Sqpy geometry and absorption energy increases as hydrogen penetrates deeper within the lattice. Regardless of the surface, single hydrogen absorption into a defect-free α-U surface is an endothermic process for all facets studied here.

### III. EFFECT OF TENSILE LOAD ON SURFACE ADSORPTION, SUBSURFACE ABSORPTION, AND DIFFUSION

Due to the substantial lattice expansion of α/β-UH$_3$ compared to α-U, there is likely a significant tensile strain of the α-U lattice as more hydrogen diffuses into bulk. In this section, we investigate the surface adsorption and subsurface absorption properties as a function of applied tensile load on an α-U surface supercell of 6-10 layer thickness (with the bottom 2 layers fixed). Convergence studies indicated that a six-layer slab with a vacuum layer of 15 Å between the two adjacent slab surfaces were found to be adequate for all relevant calculations. Tensile strain was applied to each facet by first applying an isotropic strain to the bulk followed by optimization of the ion positions. We then created a surface slab for each facet and optimized its geometry while constraining the bottom two atomic layers. Interplanar distances were then computed by selecting a given surface atom and calculating the distance of the atom directly underneath.

In order to investigate thermal activation of these dilated geometries, we have computed the pressure-volume work required to create tensile loads in the bulk by fitting volumetric strain data to the Birch-Murnaghan isothermal equation of state and computing the pressure-volume work. Our results indicate that ~0.04 eV is required for the lattice to expand up to 4%, which is likely thermally accessible under ambient conditions, and we use this value as an upper bound for our study here.

Fig. 4a shows the surface energies for three standard terrace surface, (001), (010), and (100), as well as those for two kinked facets, (012), and (102), all as a function of applied tensile strain. In addition to common low-index terrace surfaces, we have analyzed a few stepped facets (Supplemental Information Fig. S1). In realistic systems, surfaces are non-ideal, defect-rich, and contain imperfections, such as kinks, steps, or defect sites. In fact, stepped facets play a significant role in the palladium hydriding process[43], implying their potential significance in this work as well. When no strain is applied, (001) is the most stable facet relative to other surfaces. In contrast, the α-U (012) kinked facet has the highest energy and, thus, the most unstable. When symmetric tensile strain is applied, the surface energy decreases for all facets. In fact, the volumetric expansion of the lattice causes densification along the surface normal as the interlayer distance decreases with applied load, which is compensated for with increased U-U bond distances within the same layer (Table S1 in SI). The effective coordination number can serve as an additional descriptor for predicting surface chemistry and activation energies[44,45] and could be included in subsequent work.

The effect of tensile strain on H atom surface adsorption and subsurface absorption are illustrated in Fig. 4(b-d) (also Table S2 in SI). Here, we compare the lowest surface adsorption to the lowest subsurface absorption site, as well as assess the trends for hydrogen diffusion from the most stable surface site to nearby subsurface site. For all facets and tensile loads, hydrogen surface adsorption prefers to bind at the hollow 3-fold sites, excluding the (010) surface where the 4-fold adsorption sites are uniformly favored. On the other hand, the tetrahedral subsurface site is preferred for the (001), (012), and (102) surfaces, while the square-pyramidal is favored in (010) and (100) slabs. Adsorption on the stepped facets is more energetically favorable than on terrace surfaces (Fig. 4a). For example, under 0% strain, the adsorption energies for (012) and (102) are -0.71 and -0.66 eV, while the energies for (001), (010), and (100) facets are -0.57, -0.39, and -0.58 eV, respectively. Compared to the terrace surfaces studied here, the stepped regions yield some distortion of distances and angles between surface atoms (see Fig. S2 for details) that may result in higher reactivity towards hydrogen.

In addition, we observe that tensile strain uniformly results in lower hydrogen surface adsorption energy (Fig. 4a). Similar trends are observed in H/Pt(111) and H/Rh(111) systems, and have been confirmed in phenomenological thermodynamic studies,[46] where a tensile load leads to stronger hydrogen bonding to the surface of interest.[47] When elastic strain is applied, two effects promote bonding between hydrogen and nearby uranium atoms. With increased tensile strain, the distances between all atoms within a surface layer also increase, which leads to a decreased bond strength. However, counteracting this effect is the simultaneous reduction of the distance between the hydrogen adsorbate and the subsurface U atoms, which then stabilizes adsorbate/surface interactions. In addition to the adsorption energy, we also notice tensile strain has a significant effect on the subsurface absorption energy (Fig. 4b). In fact, only 2% tensile strain is required for subsurface absorption to become an exothermic process.

As shown in Fig. 4c, the energy difference ($\Delta E$) between surface adsorption and subsurface absorption energy decreases as tensile strain increases, which means that lattice expansion can facilitate the formation of the hydride. This effect is enhanced in terrace surfaces, e.g., (001), (010), (100), compared to the stepped surfaces, (012) and (102). Since this surface penetration reaction energy decreases the most for most stable (001) facet, this surface/subsurface transition could be more sensitive to tensile load of those studied here. Overall, we observe that atomic hydrogen prefers to bind at the stepped sites, but surface to subsurface diffusion could be more likely to occur from the (001) facet.

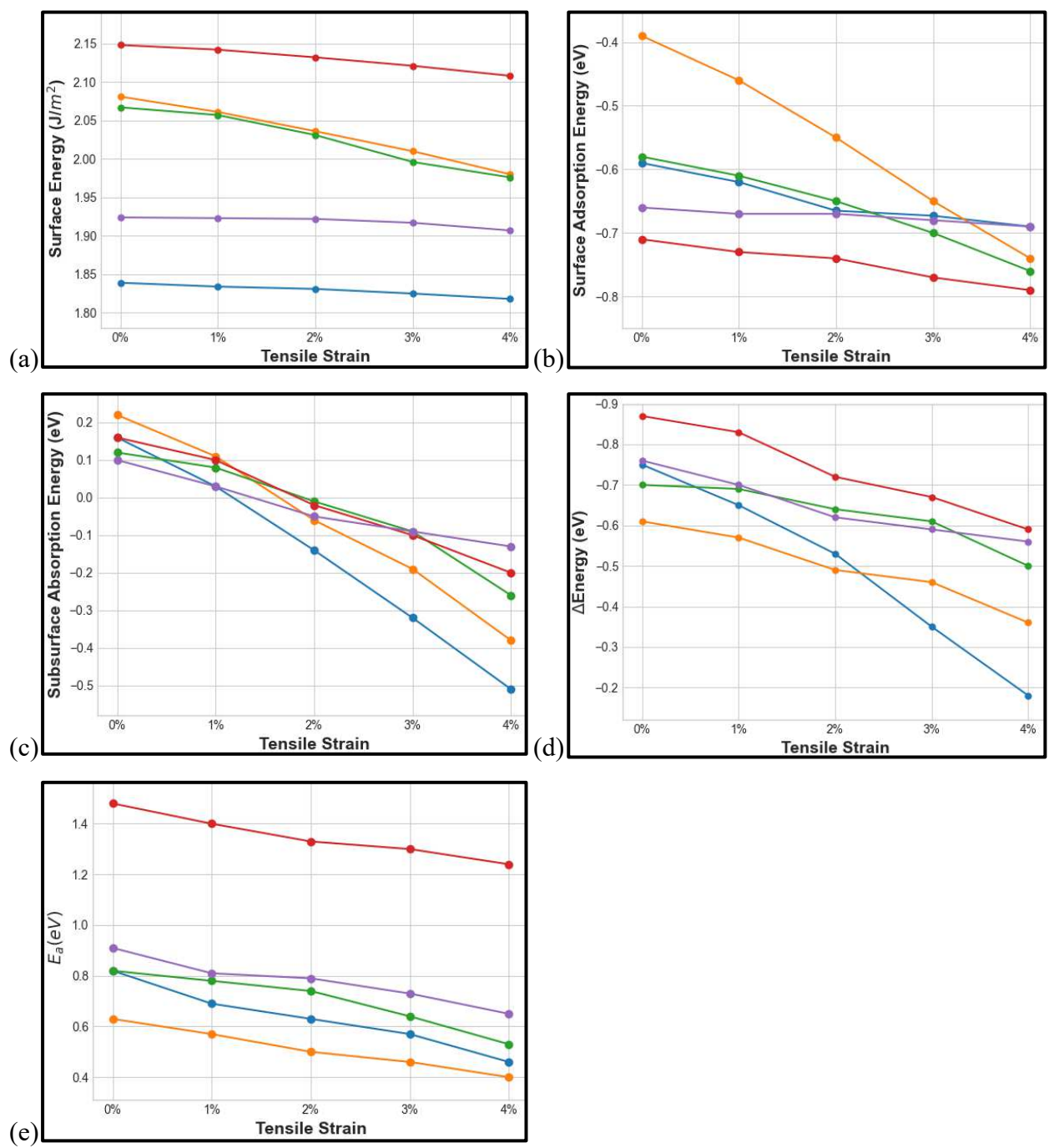

**Fig. 4:** Applied tensile strain study for (001)-blue, (010)-orange, (100)-green, (012)-red, and (102)-purple α-U facets. For a given surface and applied load: (a) the surface energy (b) the lowest adsorption energy, (c) the lowest subsurface absorption energy, (d) the energy difference (ΔE) between the surface adsorption and the subsurface absorption, and (e) the activation energy ($E_a$) for atomic hydrogen diffusion between the lowest adsorption and subsurface absorption sites for a given surface. Please see the SI for a more detailed information.

In addition to the thermodynamic properties, we have also computed activation barriers of hydrogen diffusion from the surface to first subsurface via NEB (Fig. 4e). Here, the Climbing Image Nudged Elastic Band (CI-NEB) method was employed to find the transition state and the minimum energy path (MEP) for each surface of interest. In this approach, a chain of 6-8 linear interpolation images along an initial pathway between a given initial and final states of a reaction is relaxed to determine the MEP and its corresponding saddle point. The images are relaxed along the energy pathway until the maximum residual forces on each atom are less than 0.01 eV/Å. The transition state was confirmed by presence of one imaginary vibrational frequency.

The surface to subsurface MEPs and activation energies were calculated with hydrostatic elastic tensile strain of up to 4%. Results are shown in Fig. 5a for the $\alpha$-U (001) surface (see Table S2 in the SI for results for other facets). As hydrogen penetrates the (001) surface cell, it has to overcome an activation energy barrier of 0.82 eV on the surface with no strain, 0.69 eV with 1%, 0.63 eV with 2%, 0.57 eV with 3%, and 0.46 eV with 4% tensile load. These values are in close proximity with the experimental result of 0.502 eV for activation barrier.[6] This trend is similar to our results for the adsorption and absorption energies, as well as previous calculations of hydrogen diffusion in expanded bulk palladium supercells[48]. In this case, stronger hydrogen bonding in the strained system likely also stabilizes the transition state, leading to lower activation energies overall. We also notice that the relative path from the adsorption site to the final subsurface absorption site gets shorter with tensile strain. This could be attributed to the reduction of the distance between the hydrogen adsorbate and the subsurface U atoms due to the decrease of interlayer spacing described in the above section. In addition, the plot of the activation barrier as a function of reaction energy (Fig. 5b) shows a linear dependence, while simultaneously showing an inverse relationship to the applied tensile load. Thus, this relationship can potentially be exploited as a predictive tool to estimate diffusional barriers for similar loads. As shown in Table S2, results for other surfaces follow a similar trend, with facets exhibiting lower adsorption energies (e.g., kinked vs. terrace surfaces) showing higher activation diffusion barriers overall.

In order to further probe quantum nuclear vibrational effects, we have computed the ZPE for hydrogen surface adsorption and subsurface absorption, as well as those effects on reaction energies and transition states shown in Figure 4 (see Table S2 in the SI for detailed results). We observe that the ZPE occurs as a systematic correction for all surface and subsurface adsorption sites studied here. For example, surface adsorption results in a ZPE correction of $0.15 \pm 0.02$ eV whereas as subsurface absorption results in a correction of $0.22 \pm 0.02$ eV. However, in general the ZPE effect is canceled for the reaction and transition state energies, which are very similar to their non-corrected counterparts. We also observe that for some facets, the ZPE correction results in a slight increase of the tensile load requirement for exothermic surface diffusion. However, vibrational frequencies from DFT are known to exhibit functional dependence, and choice of a different functional could affect these results[49].

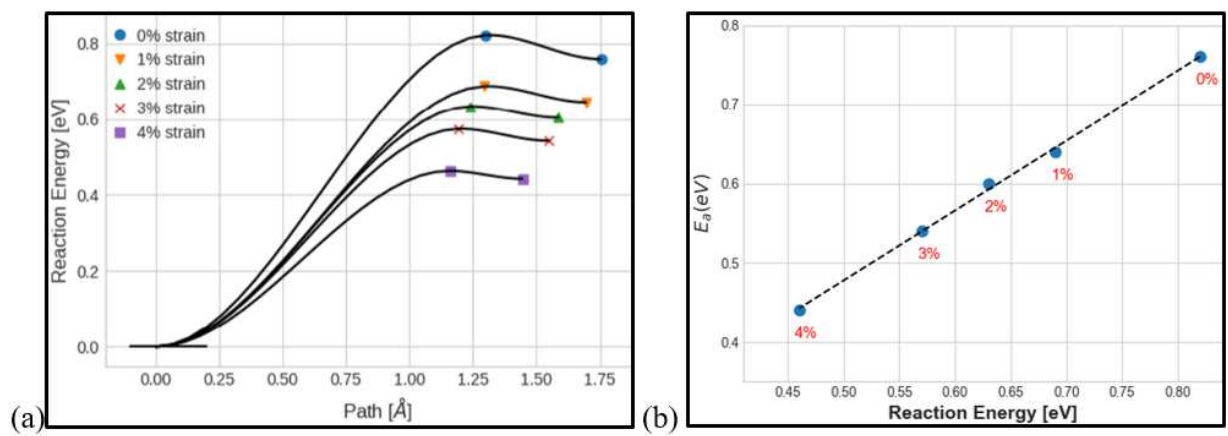

**Fig. 5**: (a) Schematic illustration of the reaction barrier for H penetration on a α-U (001) surface under tensile strain and (b) the activation energy vs reaction energy plot as a function of tensile strain.

## IV. COVERAGE DEPENDENCE ON ATOMIC HYDROGEN SURFACE ADSORPTION AND SUBSURFACE ABSORPTION

### A. Monolayer coverage dependence

Adsorption on metal surfaces is governed by both the adsorbate/surface interactions and the adsorbate/adsorbate interactions that could be attractive or repulsive. Such interactions can significantly impact the kinetics and thermodynamics of processes on surfaces, such as monolayer formation, desorption, diffusion, phase transitions, and chemical reactions.[50,51] In the previous section, all of the adsorption calculations were performed on lower surface coverage, isolated systems (e.g., in the dilute limit). However, this dilute limit does not always correspond to experimental conditions, where a range of monolayer coverages can be explored. Here, we compute properties over a range of monolayer coverages, helping to bridge the gap between DFT and experiments.

In order to understand the effects of multiple hydrogen atoms, we assess the effect of increasing hydrogen coverage on adsorption unstrained surfaces. In this case, we calculate the energetic trend for an increasing monolayer coverage by computing the adsorption energy per H atom using the following formula:

$$E_{ads} = \frac{E_{(U/H)} - E_{(U)} - 1/2\, N_H * E_{H_2}}{N_H}$$

where $E_{(U/H)}$ is the total energy of the system with adsorbates present, $E_{(U)}$ is the total energy of the clean surface, $E_{H_2}$ is the energy of gas phase hydrogen molecule, and $N_H$ is the number of adsorbed hydrogen atoms. In our study, the full monolayer (ML) coverage corresponds only to exothermic adsorption of hydrogen. For example, the (001) facet contains 20 possible adsorption sites (top, hollow, and bridge). However, only the first 8 adsorbed hydrogens in hollow sites yielded exothermic adsorption. Therefore, for (001) surface we elaborated the adsorption behavior at coverages ranging from 1/8 ML to 1 ML.

In order to overcome the combinatorial issue associated with an exhaustive search for adsorbate-surface structures, we employed a greedy forward-stepwise sampling method.[52] This approach allows for finding an adsorbate/surface configuration with a reasonably low energy using a minimal amount of computational resources. First, we determine the most favorable configuration for a single hydrogen atom on a given facet. Then, we take that configuration and add another hydrogen atom to different sites until we determine the lowest energy configuration for that particular coverage. The process is repeated step-wise until the full monolayer coverage is achieved.

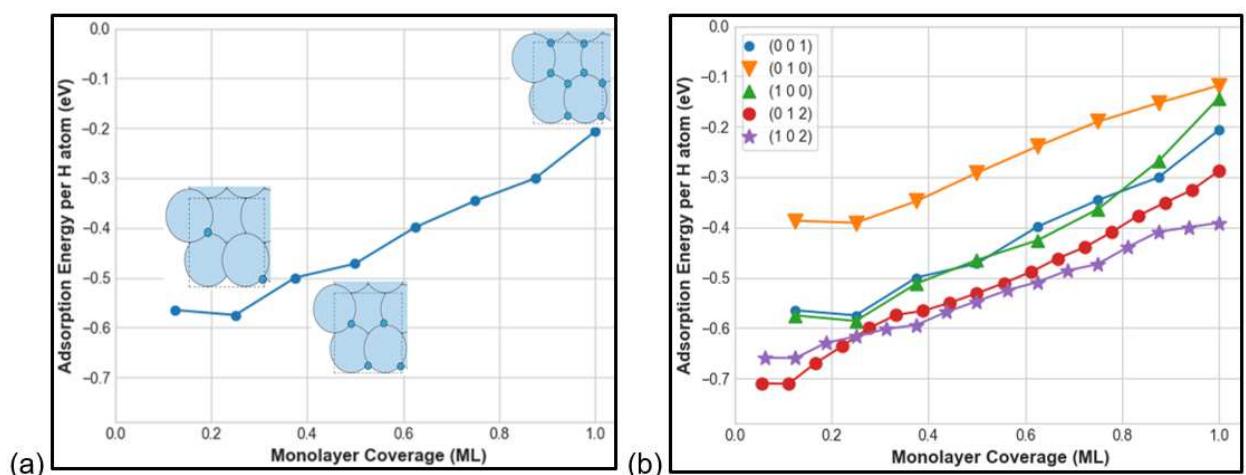

**Fig. 6**: Adsorption energy per H atom on α-U surface at different coverage: (a) (001) facet, (b) a cross-facet comparison including the (001)-blue, (010)-orange, (100)-green, (012)-red, and (102)-purple α-U surfaces.

As shown in Fig. 6, different structural combinations at each specific monolayer coverage produce a dissimilar average adsorption energy. In addition, the average adsorption energy becomes more positive as more hydrogen adsorbs on any given surface. For the (001) surface (Fig. 6a), the first H adatom prefers to bind to the hollow2 site, as previously described. At 0.25 ML coverage, the second hydrogen atom prefers to bind nearest hollow2 site. At this point, we observe some electrostatic screening from uranium atoms (flat line) as there are little to no interactions between two adsorbed hydrogen atoms that are spatially close to each other. In fact, the same preference occurs up until all hollow2 sites are occupied. At 0.5 ML coverage, the four hydrogen adatoms form a period zigzag structure on the α-U (001) surface. Newly added H atoms adsorb to nearby available hollow1 sites until all hollow sites are occupied at 1 ML coverage.

We observe similar trends on other terrace and stepped facets studied here (Fig. 6b). At low coverage, we observe likely screening effects (flat line) in all surfaces, similar to the (001) case. In addition, the adsorption energy per hydrogen atom becomes more positive (repulsive) with increasing monolayer coverage. In terms of the energy scale, the kinked facets (012) and (102) start at lower energies (-0.71 eV/H and -0.66 eV/H, respectively) compared to terrace surfaces, as they possess the most reactive stepped sites. As coverage increases, the stepped sites on (012) and (102) fill up first, followed by adsorption at hollow sites in relatively (unkinked) flat regions of the surface. In contrast, in (010) and (100), we observe initial adsorption at hollow sites followed by adsorption into bridge sites. For all surfaces, no exothermic adsorption occurs on top sites directly above U surface atoms.

### B. Surface H/U phase diagram

All of our previous calculations were based on zero-temperature and zero-pressure optimizations. In order to connect our DFT calculations to actual thermodynamic measurements, we use the Python multiscale thermochemistry toolbox[53] (pMuTT) to generate the surface phase diagram for equilibrium hydrogen adsorption. pMuTT is a Python library designed to estimate various thermochemical and kinetic properties from ab-initio data for heterogeneous catalysis. Here, the 2D phase diagram is computed by finding the U/H surface configuration from our set of calculations that exhibits the lowest Gibbs free energy at a specific temperature and pressure. The Gibbs free energy for each adsorbate-solid structure is estimated in the harmonic limit, where each adsorbed hydrogen is treated as an independent quantum-harmonic oscillator and its vibrational frequencies are computed via finite difference.

In Fig. 7, each computed phase diagram depicts the monolayer coverage as a function of pressure and temperature. Configurations with high monolayer coverage are more favored at high pressure and low temperature, whereas the clean α-U surface is more prominent at low pressure and high temperature. As temperature increases, the entropy of the gas phase drives the desorption of hydrogen. Since hydrogen adsorption on an α-U surface is an exothermic reaction, the adsorption capacity increases with the decreasing temperature. We also notice that only the (102) facet shows a complete monolayer coverage at low temperatures (~150 K) and high pressure (>1 bar) over the range of conditions studied here. On the other hand, facets behave differently at various non-extreme cases. If we look at ambient conditions of 300 K and a realistic partial pressure of hydrogen gas in air (~$10^{-4}$ bar), these stepped surfaces have higher monolayer coverages than the terrace facets. Specifically, the (012) and (102) facets have 0.56 ML coverage at these conditions, while (001), (100), and (010) have smaller monolayer coverages of 0.38 ML, 0.5 ML, and 0.25 ML, respectively. At 600 K, which is the operating temperature some nuclear reactors, and at ~$10^{-4}$ bar hydrogen partial pressure, we notice a similar trend. We note that this temperature is above the transition point for α to β-UH$_3$ under atmospheric pressure. The stepped facets have a monolayer coverage of 0.17 ML, while the terrace surfaces have 0 ML coverage. Overall, we observe that monolayer coverage is generally more significant for stepped surfaces over all conditions studied here, consistent with the conventional notion that these surfaces are likely much more reactive than their terrace counterparts.

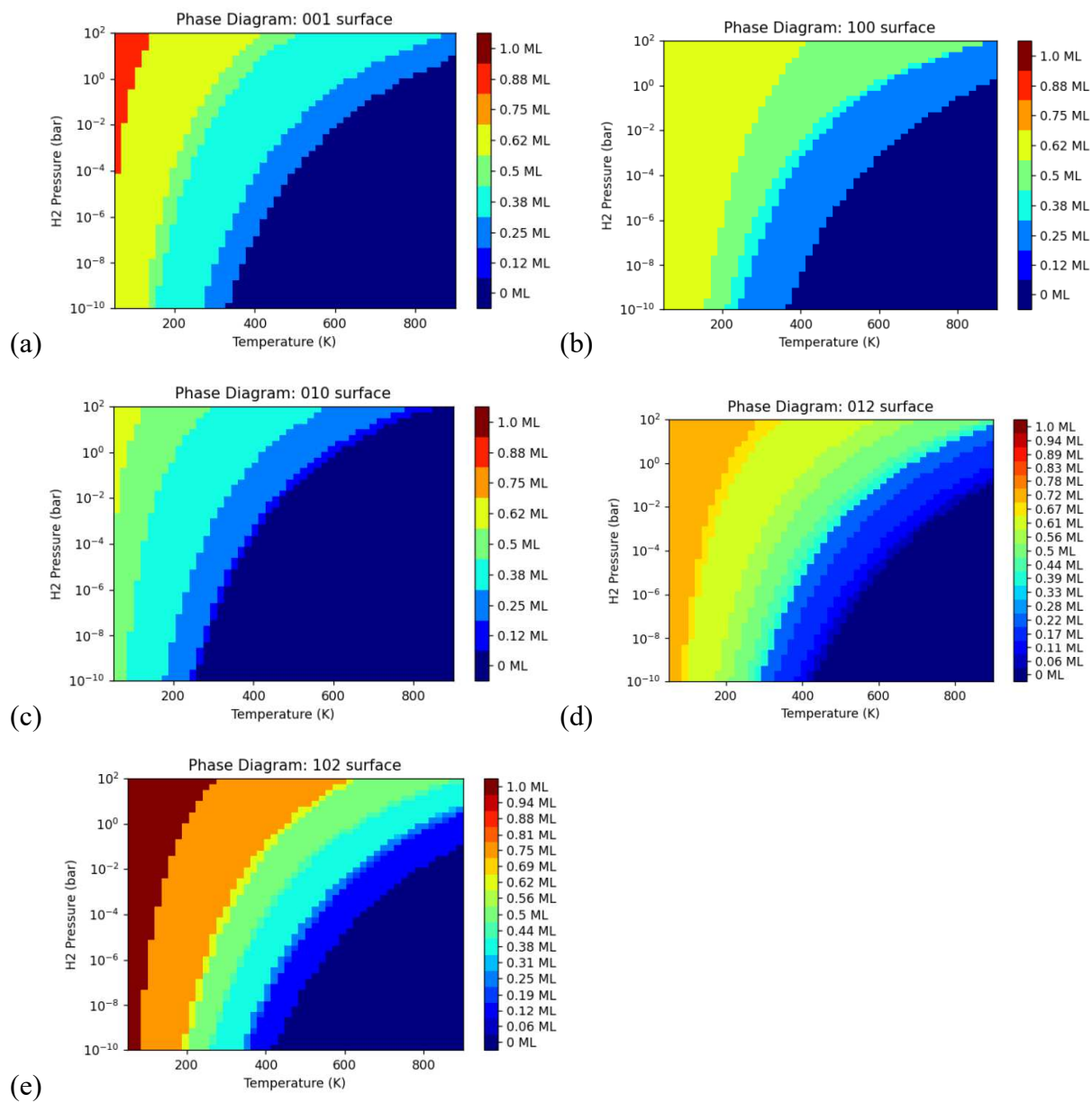

**Fig. 7**: Ab-initio phase diagram of the H/α-U system generated by pMuTT, where the color represents the most stable configuration at a given temperature and pressure (a) (001), (b) (010), (c) (100), (d) (012), and (e) (102).

## C. Subsurface adsorption as a function of surface coverage

We now investigate the possibility of surface adsorption either mitigating or enhancing subsurface absorption for a non-tensile loaded system. Our goals are to potentially find a cross-over point where hydrogen subsurface absorption becomes more favorable compared to surface adsorption for various facets of interest. We compute the energy to add hydrogen into either the first subsurface layer or on the surface itself for each ML coverage using the following equation:

$$E_{add_surf/add_subsurf} = E_{(U/NH+1)} - E_{(U/NH)} - \frac{1}{2} * E_{H_2}$$

Here, $E_{(U/NH+1)}$ is the energy of the system with N+1 surface adsorbates (for $E_{add\ surf}$) or the energy of the system with N surface adsorbates and one subsurface hydrogen (for $E_{add\ subsurf}$), $E_{(U/NH)}$ is the total energy of the system with N surface adsorbates, and $N$ is the number of adsorbed hydrogen atoms.

Table 6 shows the results for surface coverage of the (001) facet. At zero ML, adding an H atom to the surface is favored over the subsurface; in fact, subsurface adsorption is endothermic (+0.16 eV), while surface binding is exothermic (-0.57 eV). The most stable subsurface site at this point is used in subsequent calculations at higher coverages. Subsurface adsorption becomes exothermic at 1/8 ML coverage, though adding hydrogen to the surface is still more favored. This trend continues until 3/8 ML coverage, where the subsurface absorption (-0.36 eV) becomes preferred over surface absorption (-0.26 eV). Further details on the other low index terrace and kinked surfaces and the results are presented in the Supplemental Information in Table S3. For all other facets studied here, subsurface absorption was either endothermic or still less favorable compared to the surface adsorption. As a result, (001) is the most likely candidate from our specific set that appears able to initiate subsurface hydriding at higher ML coverage.

TABLE 6: The adsorption energy to add a hydrogen atom on (2x1) α-U (001) surface or in the first subsurface layer as a function of coverage. $E_{add_{subsurf}}$ denotes the energy required to add a hydrogen atom to the subsurface layer. $E_{add_{surf}}$ (eV) represents the energy required to add a hydrogen on the top of the surface.

| Surface Coverage | $E_{add\_subsurf}$ (eV) | $E_{add\_surf}$ (eV) |
|---|---|---|
| 0 | 0.16 | -0.57 |
| 1/8 (ML) | -0.02 | -0.44 |
| 1/4 (ML) | -0.11 | -0.43 |
| 3/8 (ML) | -0.36 | -0.26 |
| 1/2 (ML) | -0.44 | -0.21 |

## CONCLUSION

In this work, we have investigated uranium hydriding from two perspectives: as a function of (1) applied tensile load on a α-U lattice and (2) formation of hydrogen monolayer coverage. First, we have analyzed the adsorption, absorption, and diffusion of hydrogen from surface to the first subsurface layer under applied symmetric tensile strain of up to 4%. Absorption sites closest to the surface were found to be the most favorable and exhibited either tetrahedral or square-pyramidal geometries. Both surface and subsurface binding energies were found to be sensitive to the hydrostatic elastic tensile strain, leading to lower values overall. In fact, a tensile strain of only 2% is required for subsurface absorption to become an exothermic process, though this value could shift somewhat depending on the choice of DFT functional and inclusion zero-point energy effects. In addition, the energy difference between surface adsorption and first layer subsurface absorption becomes less positive with increasing tensile strain. These tensile strain states are likely thermally accessible at ambient conditions. Similar to the binding energies, the diffusion barriers were found to be responsive to strain and exhibited an inverse linear relationship as a function of reaction energy, yielding a new predictive capability for this process.

We also explore the hydriding process from the perspective of formation of a hydrogen monolayer on the uranium surface. Coverage studies for all facets showed that the adsorption energy becomes more positive as more atoms hydrogen adsorb on the surface. We also observe that the kinked (012) and (102) facets initially exhibit more exothermic adsorption energies compared to the terrace surfaces. As depicted in our phase diagrams, they also show larger monolayer coverages with increasing pressure and temperature. Also, calculation of the energetic cost of inserting additional hydrogen inside the first subsurface layer of the (001) facet showed: (1) at 1/8 ML, subsurface absorption becomes exothermic, and (2) at 3/8 ML, this absorption is energetically favored over surface adsorption. This trend was only observed for the (001) facet in our study. Hence, we have found that high monolayer coverage is more likely on the (102) surface, with 56% coverage possible at ambient conditions. In contrast, (001) ML formation is more likely to enhance subsurface penetration. Hydrogen diffusion kinetics from a metallic surfaces to its subsurface is strongly affected by monolayer coverage[54], which is the subject of future investigations. The types of results presented in our work can be important inputs for larger scale models, such as kinetic Monte Carlo, that can further bridge the time and length scale gap with experiments. This could include looking at defect systems, larger scale models that can bridge gaps with experiments, and varied atmospheric conditions.

# Elucidating the initial steps in α–uranium hydriding using first-principles calculations

# Supplemental Information

*By Artem Soshnikov,[1] Ambarish Kulkarni,[1] Nir Goldman[1,2]*

[1]Department of Chemical Engineering, University of California, Davis, CA 95616, USA

[2]Physical and Life Sciences Directorate, Lawrence Livermore National Laboratory, Livermore, CA, 94550 USA.

Email: goldman14@llnl.gov

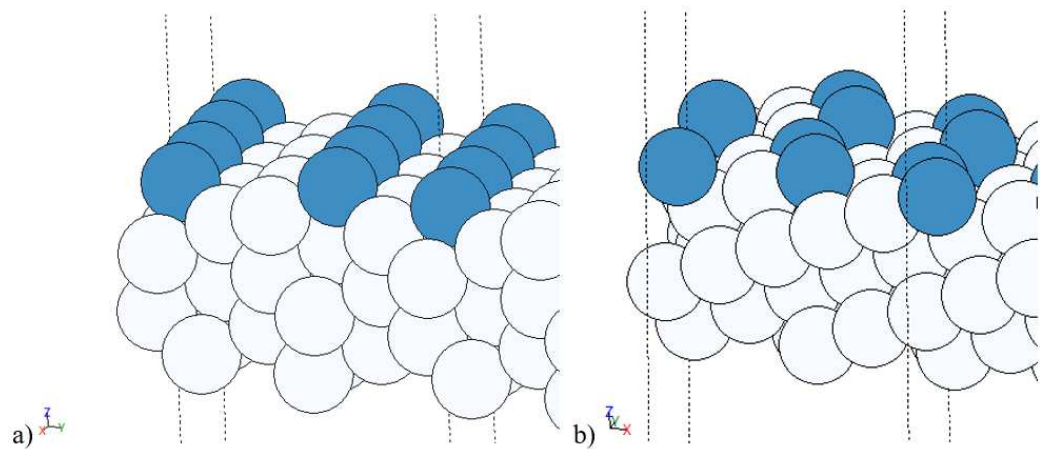

**Fig. S1**: α-U kinked surfaces. Kinked U atoms are highlighted in blue. a) (012), b) (102).

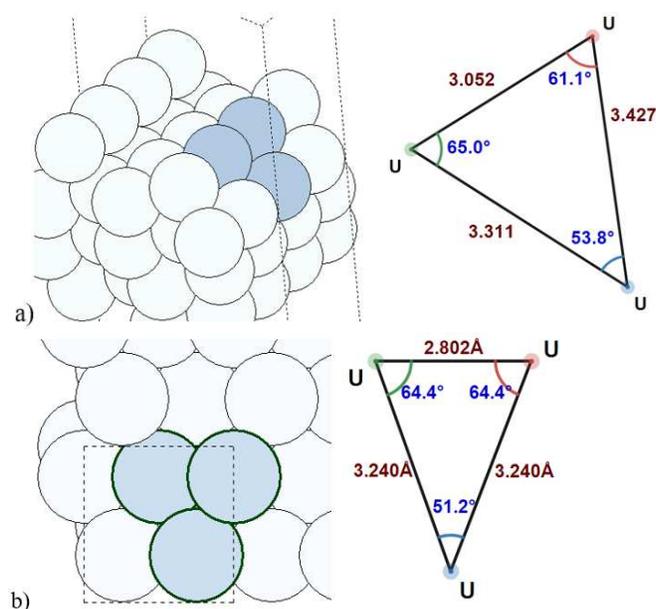

**Fig. S2:** Geometric comparison of terrace and kinked sites. (a) stepped hollow site of (012) facet, (b) terrace hollow site of (001) surface. (012) facet has 3.052, 3.311, and 3.196 Å U-U bond distances and 53.8°, 61.1°, and 65.0° angles that encompass the most energetically favorable hollow site, compared to values of 2.802 Å and 3.240 Å for bonds and 51.2°, 64.4°, and 64.4° for angles from the (001) surface.

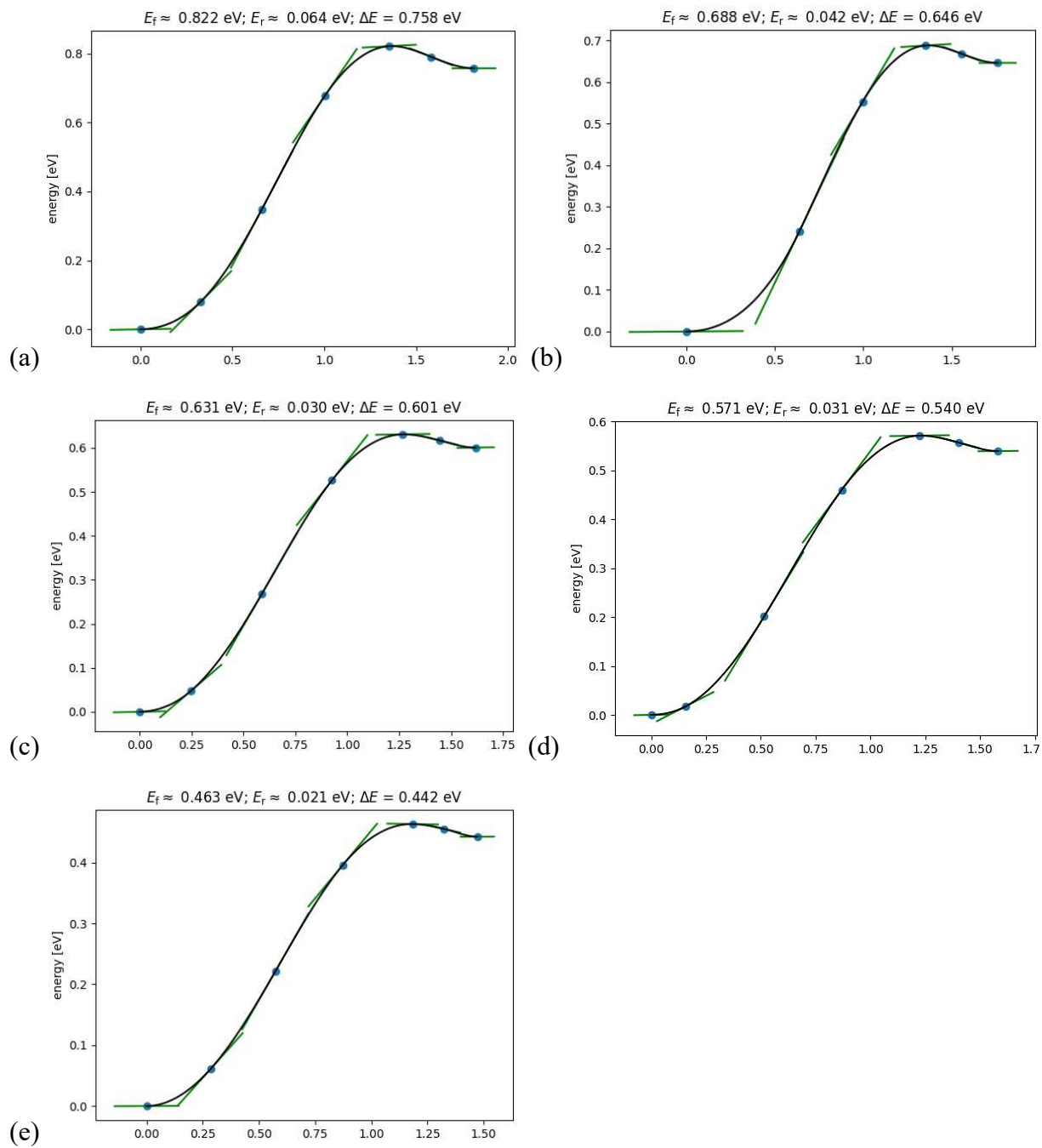

**Fig S3.** Sample NEB plots for (001) facet. (a) 0%, (b) 1%, (c) 2%, (d) 3%, and (e) 4% tensile load.

**TABLE S1:** Interlayer and U-U bond distances comparison as a function of tensile strain for common terrace surfaces.

| Surface | Expansion | Interlayer distance (Å) | % change due to exp. (rel. to perf) | % change due relax. (rel. to perf.) | % total change (rel. to perf. lattice) | Bond distance (Å) | % change (rel. to perf. lattice) |
|---|---|---|---|---|---|---|---|
| 001 | 0% | 2.353 | 0.00% | 0.00% | 0.00% | 2.802 | 0.00% |
|  | 1% | 2.338 | 1.00% | -1.64% | -0.64% | 2.83 | 1.00% |
|  | 2% | 2.324 | 2.00% | -3.23% | -1.23% | 2.858 | 2.00% |
|  | 3% | 2.311 | 3.00% | -4.78% | -1.78% | 2.886 | 3.00% |
|  | 4% | 2.297 | 4.00% | -6.38% | -2.38% | 2.914 | 4.00% |
| 010 | 0% | 2.903 | 0.00% | 0.00% | 0.00% | 2.802 | 0.00% |
|  | 1% | 2.887 | 1.00% | -1.55% | -0.55% | 2.83 | 1.00% |
|  | 2% | 2.869 | 2.00% | -3.17% | -1.17% | 2.858 | 2.00% |
|  | 3% | 2.851 | 3.00% | -4.79% | -1.79% | 2.886 | 3.00% |
|  | 4% | 2.833 | 4.00% | -6.41% | -2.41% | 2.914 | 4.00% |
| 100 | 0% | 2.624 | 0.00% | 0.00% | 0.00% | 2.732 | 0.00% |
|  | 1% | 2.608 | 1.00% | -1.61% | -0.61% | 2.751 | 0.70% |
|  | 2% | 2.596 | 2.00% | -3.07% | -1.07% | 2.767 | 1.28% |
|  | 3% | 2.583 | 3.00% | -4.56% | -1.56% | 2.786 | 1.98% |
|  | 4% | 2.572 | 4.00% | -5.98% | -1.98% | 2.807 | 2.75% |

**TABLE S2:** Surface vs subsurface adsorption energy comparison as a function of tensile strain. $E_{ads}$ represents the lowest adsorption energy for H atom on surface; $E_{ss}$ denotes the lowest absorption energy to insert H atom inside the first subsurface layer for a given surface; $\Delta E$ indicates the energy difference between surface adsorption and subsurface absorption; $E_{TS}$ represents the energy of the transition state.

| Surface | Tensile strain | Surf site | $E_{ads}$ (eV) | $E_{ads}$ (eV) w/ZPE | Subsurf site | $E_{ss}$ (eV) | $E_{ss}$ (eV) w/ZPE | $\Delta E$ (eV) | $E_{TS}$ (eV) | $E_{TS}$ (eV) w/ZPE |
|---|---|---|---|---|---|---|---|---|---|---|
| 001 | 0% | 3-fold | -0.57 | -0.41 | Tet | 0.16 | 0.40 | -0.73 | 0.82 | 0.86 |
| | 1% | 3-fold | -0.62 | -0.46 | Tet | 0.03 | 0.26 | -0.65 | 0.69 | 0.73 |
| | 2% | 3-fold | -0.67 | -0.52 | Tet | -0.14 | 0.09 | -0.53 | 0.63 | 0.66 |
| | 3% | 3-fold | -0.67 | -0.52 | Tet | -0.32 | -0.10 | -0.35 | 0.57 | 0.60 |
| | 4% | 3-fold | -0.69 | -0.53 | Tet | -0.51 | -0.30 | -0.18 | 0.46 | 0.48 |
| 010 | 0% | 4-fold | -0.39 | -0.25 | Sqpy | 0.22 | 0.43 | -0.61 | 0.78 | 0.82 |
| | 1% | 4-fold | -0.46 | -0.32 | Sqpy | 0.11 | 0.32 | -0.57 | 0.67 | 0.71 |
| | 2% | 4-fold | -0.55 | -0.41 | Sqpy | -0.06 | 0.15 | -0.49 | 0.55 | 0.59 |
| | 3% | 4-fold | -0.65 | -0.52 | Sqpy | -0.19 | 0.01 | -0.46 | 0.49 | 0.52 |
| | 4% | 4-fold | -0.74 | -0.61 | Sqpy | -0.38 | -0.20 | -0.36 | 0.42 | 0.45 |
| 100 | 0% | 3-fold | -0.58 | -0.42 | Sqpy | 0.12 | 0.33 | -0.70 | 0.82 | 0.83 |
| | 1% | 3-fold | -0.61 | -0.44 | Sqpy | 0.08 | 0.29 | -0.69 | 0.78 | 0.78 |
| | 2% | 3-fold | -0.65 | -0.49 | Sqpy | -0.01 | 0.19 | -0.64 | 0.74 | 0.73 |
| | 3% | 3-fold | -0.70 | -0.54 | Sqpy | -0.09 | 0.11 | -0.61 | 0.64 | 0.66 |
| | 4% | 3-fold | -0.76 | -0.60 | Sqpy | -0.26 | -0.07 | -0.50 | 0.55 | 0.53 |
| 012 | 0% | 3-fold | -0.71 | -0.54 | Tet | 0.16 | 0.39 | -0.87 | 1.48 | 1.49 |
| | 1% | 3-fold | -0.73 | -0.56 | Tet | 0.10 | 0.33 | -0.83 | 1.40 | 1.42 |
| | 2% | 3-fold | -0.74 | -0.57 | Tet | -0.02 | 0.20 | -0.72 | 1.33 | 1.33 |
| | 3% | 3-fold | -0.77 | -0.60 | Tet | -0.10 | 0.12 | -0.67 | 1.30 | 1.30 |
| | 4% | 3-fold | -0.79 | -0.62 | Tet | -0.20 | -0.01 | -0.59 | 1.24 | 1.25 |
| 102 | 0% | 3-fold | -0.66 | -0.52 | Tet | 0.10 | 0.33 | -0.76 | 0.91 | 0.95 |
| | 1% | 3-fold | -0.67 | -0.53 | Tet | 0.03 | 0.25 | -0.70 | 0.81 | 0.85 |
| | 2% | 3-fold | -0.67 | -0.53 | Tet | -0.05 | 0.14 | -0.62 | 0.79 | 0.83 |
| | 3% | 3-fold | -0.68 | -0.54 | Tet | -0.09 | 0.10 | -0.59 | 0.73 | 0.74 |
| | 4% | 3-fold | -0.69 | -0.55 | Tet | -0.13 | 0.05 | -0.56 | 0.65 | 0.66 |

**TABLE S3**: The adsorption energy to add additional hydrogen atom on α-U surface or in the first subsurface layer as a function of coverage. $E_{add_{subsurf}}$ denotes the energy required to add a hydrogen atom to the subsurface layer; $E_{add_{surf}}$ (eV) represents the energy required to add a hydrogen on the top of the surface.

| Surface | Coverage (ML) | $E_{add\_surf}$ (eV) | $E_{add\_subsurf}$ (eV) |
|---|---|---|---|
| (010) | 0 | -0.39 | 0.22 |
| | 1/8 | -0.40 | 0.24 |
| | 2/8 | -0.26 | 0.25 |
| | 3/8 | -0.12 | 0.25 |
| | 4/8 | -0.03 | 0.22 |
| | 5/8 | 0.06 | 0.32 |
| (100) | 0 | -0.58 | 0.12 |
| | 1/8 | -0.60 | 0.20 |
| | 2/8 | -0.37 | 0.19 |
| | 3/8 | -0.32 | 0.17 |
| | 4/8 | -0.28 | 0.18 |
| | 5/8 | -0.11 | 0.25 |
| | 6/8 | 0.26 | 0.28 |
| (012) | 0 | -0.71 | 0.16 |
| | 1/18 | -0.67 | 0.17 |
| | 2/18 | -0.63 | 0.18 |
| | 3/18 | -0.53 | 0.19 |
| | 4/18 | -0.43 | 0.17 |
| | 5/18 | -0.41 | 0.10 |
| | 6/18 | -0.40 | 0.06 |
| | 7/18 | -0.38 | 0.05 |
| | 8/18 | -0.38 | 0.11 |
| | 9/18 | -0.33 | 0.12 |
| (102) | 0 | -0.66 | 0.10 |
| | 1/16 | -0.66 | -0.07 |
| | 2/16 | -0.57 | -0.09 |
| | 3/16 | -0.54 | -0.09 |
| | 4/16 | -0.56 | -0.14 |
| | 5/16 | -0.41 | -0.15 |
| | 6/16 | -0.40 | -0.10 |
| | 7/16 | -0.35 | -0.08 |
| | 8/16 | -0.34 | 0.21 |
| | 9/16 | -0.26 | 0.27 |